\documentclass{vgtc}                          

\graphicspath{{figures/}{pictures/}{images/}{./}} 

\usepackage{times}                     

\usepackage{tabu}                      
\usepackage{booktabs}                  
\usepackage{lipsum}                    
\usepackage{mwe}                       
\usepackage{makecell}
\usepackage{enumitem}
\usepackage{cite}

\usepackage{mathptmx}                  

\onlineid{1052}

\vgtccategory{Research}

\vgtcinsertpkg

\newif\ifshowcomments
\showcommentsfalse

\newcommand{\revised}[1]{\ifshowcomments{\color{blue}{#1}}\else{#1}\fi}

\newcommand{\cut}[1]{\ifshowcomments{\color{red}{#1}}\else\fi}

\title{How Do Observable Users Decompose D3 Code? A Qualitative
Study}

\author{Melissa Lin\thanks{Both authors contributed equally to this research}~~\thanks{e-mail: mylin@andrew.cmu.edu}\\ %
    \scriptsize Carnegie Mellon University %
\and Heer Patel\footnotemark[1]~~\thanks{e-mail: [heerpate, mlamkin, leibatt]@cs.washington.edu}\\ %
    \scriptsize University of Washington %
\and Medina Lamkin \footnotemark[3]\\ 
    \scriptsize University of Washington \\ 
\and Hannah Bako \thanks{e-mail: hbako@umd.edu} \\
    \scriptsize University of Maryland \\
\and Leilani Battle \footnotemark[3] \\ 
    \scriptsize University of Washington \\
\vspace{-15mm}
}
\abstract{
    Many toolkit developers seek to streamline the visualization programming process \cut{for their users }through structured support such as prescribed templates and example galleries. However, few projects examine how users organize their own visualization programs \revised{and} how their coding choices may deviate from the intents of toolkit developers, \cut{and how these differences may }impacting visualization prototyping and design. Further, is it possible to infer users' reasoning indirectly through their code, even when users copy code from other sources? \cut{Understanding these patterns can reveal opportunities to align toolkit design with actual user behavior, improving usability and supporting more flexible workflows.} We explore this question through a qualitative analysis of 715 D3 programs on Observable. We identify three levels of program organization based on how users decompose their code into smaller blocks: Program-, Chart-, and Component-Level code decomposition, with a strong preference for Component-Level reasoning. In a series of interviews, we corroborate that these levels reflect how Observable users reason about visualization programs. We compare common user-made components with those theorized in the Grammar of Graphics to assess overlap in user and toolkit developer reasoning. We find that, while the Grammar of Graphics covers basic visualizations well, it falls short in describing complex visualization types, especially those with animation, interaction, and parameterization components. Our findings highlight how user practices differ from formal grammars and 
\revised{reinforce ongoing efforts to rethink}
visualization toolkit support, including augmenting learning tools and AI assistants to better reflect real-world coding strategies.
} 


\begin{document}


\maketitle


\section{Introduction}
\label{sec:introduction}

From designing bespoke visualizations in D3~\cite{Bostock2011d3} to orchestrating multi-chart interactions in Vega-Lite~\cite{Satyanarayan2017vegalite}, visualization programming is a valuable skill for the data science workforce~\cite{ryan2019teaching}.
However, users may struggle to write their own customized programs~\cite{Bako2023userdriven}, even when adapting existing examples~\cite{Battle2022exploring}, making it challenging to adopt more expressive toolkits like D3~\cite{Satyanarayan2020critical}. One popular solution is to generate code templates~\cite{McNutt2021integrated} that users can populate with data variables~\cite{Bako2022streamlining, Harper2018converting}. However, templates do not necessarily reflect how users themselves may reason about their own code~\cite{Bako2023userdriven}. Broadly, we observe relatively few research projects investigating how visualization toolkit users organize their code, or how these choices impact code comprehension and reuse~\cite{Battle2022exploring,Pu2023howdata}.
We focus on D3 given its popularity and complexity~\cite{Battle2022exploring,Bako2023userdriven,Harper2018converting}.

In response, we analyze how D3 users organize visualization code to understand what code structures these users find intuitive.
\cut{Specifically,} We apply qualitative methods to investigate an established measurement of code comprehension \cut{from the CS education literature}: code decomposition~\cite{rich2018decomposition,charitsis2023detecting, tang2020assessing}.
\revised{The abstraction and pattern-matching skills necessary for code decomposition make it a critical pillar of \emph{computational thinking} that helps to distinguish novice programmers from experts~\cite{yadav2022computational,tang2020assessing}. Further, practicing program decomposition helps students develop \emph{metacognitive awareness}, and researchers have analyzed decomposition strategies to observe students' mental models of code in the classroom~\cite{charitsis2023detecting}.}
\revised{With visualization languages, users must reason about their code \emph{and} visualization designs simultaneously~\cite{Bako2023userdriven,Battle2022exploring,Pu2023howdata}, suggesting that decomposition strategies could be analyzed to gain insight into users' mental models of visualization programs.}
Through this analysis, we can establish frameworks for users’ mental models to ground visualization programming and design tools in actual user behavior.
Towards this goal, we address two core research questions: \emph{How do D3 users (re)organize code copied from outside sources} (\textbf{RQ1})? Further, \emph{do D3 users really organize their code according to toolkit designers' recommendations} (\textbf{RQ2})?

To answer these questions,
we contribute a qualitative analysis of D3 code decomposition strategies across 715 Observable~\cite{Bostock2016observable} notebooks representing 24 distinct visualization types.
Given D3 users often copy from existing examples~\cite{Battle2022exploring,Harper2018converting, Hoque2020searching,McNutt2021integrated}, 
we focus our first analysis on how Observable users organize copied code (answering \textbf{RQ1}).
Our findings reveal three distinct granularities for decomposing D3 programs into smaller code blocks: component, chart, and program level. Component-Level decomposition was the most prevalent strategy.
We complement our analysis with interviews of 7 Observable users, who confirm that they are purposeful in how they structure their D3 code and share how their code structuring is influenced by (inferred) toolkit and community best practices.

To understand how users' mental models align with existing toolkit design paradigms (answering \textbf{RQ2}), we compare common user-made D3 components with the Grammar of Graphics (GoG). We find high component overlap for common chart types such as bar \cut{and line }charts but diminishing coverage for more complex \revised{ones} such as \cut{treemaps and }streamgraphs. Further, several \revised{commonly used} D3 component types, such as interaction and parameterization, fall outside \cut{the scope of }the GoG, suggesting gaps between theory and real-world usage. These \revised{user-driven} insights \revised{validate ongoing efforts in visualization grammar development and reveal new} opportunities for building D3 support tools and \cut{educational }resources that better reflect users' code organization strategies. In summary, this paper makes \revised{four} contributions:
 \begin{itemize}[nosep]
    \item A qualitative analysis of 715 Observable notebooks identifying three levels of code decomposition and the impacts of code copying on decomposition strategies.
    \item An interview study with 7 Observable users, which corroborates our qualitative findings and clarifies user rationales for adopting different decomposition strategies.
    \item A comparison of user-made D3 code components with the GoG, revealing actionable gaps between theory and practice.
    \item Design implications for tailoring educational materials and AI-driven support to match real-word toolkit usage.
\end{itemize}

\section{Related Work}
\label{sec:related-work}

\textbf{Visualization Code Reuse:} Copying from examples is a key user strategy for creating D3 visualizations ~\cite{Battle2022exploring,Bako2023understanding,Bako2023userdriven}. For example, in their analysis of 37,815 D3-related posts on Stack Overflow, Battle et al. observe that 14\% of them reference just three sources: Observable, the D3 gallery, or Bl.ocks.org ~\cite{Battle2022exploring}. However, \emph{D3 code often contains uncommon syntax and code structures, making it difficult to reuse}~\cite{Battle2022exploring}. In-depth examples may over-complicate prototyping~\cite{Battle2022exploring} or even lead to design fixation~\cite{Parsons2021fixation}.
\cut{We seek to understand how users structure their D3 code, which can further clarify how visualization paradigms support user programming in practice.}
Example galleries \revised{also facilitate}\cut{are another dominant support for} reuse. Yang et al. find that while users want larger galleries, creators struggle to maintain them~\cite{yang2024considering}. Code reuse also has pedagogical value. Recent work showed students who adapted D3 examples created more bespoke visualizations and improved their understanding of the code~\cite{hedayati2023choose}. 
\revised{Our work complements these findings by studying how example reuse impacts code organization, revealing the impact of galleries on coding structure.}
\cut{This work highlights inherent challenges of adapting D3 code, reinforcing the need to understand how users reason about code in order to better support them.}

\textbf{Visualization Templates:} \cut{Code reuse is also common in visualization toolkits, which }Several projects aim to support \revised{code reuse} through \emph{code templates}. For example, Bako et al.
contribute templates for common D3 visualization and interaction types~\cite{Bako2022streamlining}. Harper et al. propose techniques for converting existing D3 visualizations into templates~\cite{Harper2018converting}. Tools such as Ivy generalize these ideas to aid creation and reuse~\cite{McNutt2021integrated}. However, templates are difficult to modify beyond their defined parameters, potentially impeding users' creativity and workflows~\cite{Bako2023userdriven}.
Recent work has introduced structured methods for evaluating toolkit notations,
reflecting a growing need to systematize how users and developers articulate data transformation and visual mapping \cite{kruchten2024metrics}.
\revised{Similarly, we aim to strengthen the connection between toolkit usage and design.}
\cut{Similarly, our research aims to draw more direct comparisons between users' and toolkit designers' organization of visualization programs.}

\textbf{Code Decomposition and Visualization Grammars:} Code decomposition is \revised{an established} measure of \revised{student} comprehension \revised{and reasoning} in CS education~\cite{tang2020assessing,rich2018decomposition,yadav2022computational}. \revised{For example, analyzing program decomposition strategies can elucidate relationships between students' metacognitive awareness and assignment outcomes~\cite{charitsis2023detecting}.}
\revised{Outside the classroom, notebook decomposition strategies have been analyzed to measure how users reason about data science workflows~\cite{rule2018exploration,raghunandan2023code} and automatically reorganize computational notebooks to improve clarity~\cite{Titov2022ReSplit}.}
\revised{Recent work also studies recurring patterns in D3 program structures across GitHub, Bl.ocks.org, and Observable~\cite{Bako2023userdriven,Bako2022streamlining,Harper2018converting}. We adopt a similar analysis approach. Note that while Observable's cell structure makes it easier to decompose code, researchers have been observing similar D3 program decomposition prior to Observable as well~\cite{Battle2022exploring,Bako2022streamlining}.}
\cut{These findings align with our study, where users shift decomposition strategies to support iterative thinking.}

\cut{While} Formalisms such as the Grammar of Graphics~\cite{Wilkinson2005graph} and Layered Grammar of Graphics~\cite{Wickham2010layered}
\revised{have guided} toolkit design \revised{for many years and continue to influence recent work (e.g., \cite{kay2024ggdist, kim2024erie, batziakoudi2025lost}}).
\revised{However}, \emph{it is unclear whether they align with how users reason about code.} One study found that users often reuse examples based on layout or task similarity, factors not captured by formal grammars \cite{bako2024unveiling}. Analyzing code structure offers a way to bridge this gap~\cite{Pu2023howdata}. \cut{We take a similar approach, focusing on D3 where users still face challenges in writing and debugging visualization programs~\cite{Battle2022exploring}.}




\section{Data Collection \& Preparation}
\label{methods}
\cut{To facilitate a rigorous qualitative analysis on how users organize their D3 code,}
We collected a diverse range of 240 Observable~\cite{Bostock2016observable} notebooks spanning 24 visualization types identified in previous work~\cite{Battle2018beagle}. We followed a strategy of searching by visualization type and screening notebooks for quality and uniqueness, detailed in supplemental materials (\autoref{sec:supplemental_materials}).
Similar to prior work~\cite{Battle2022exploring,Bako2022streamlining,Bako2023understanding}, we observe that most Observable notebooks are \emph{duplicates} that copy code from older notebooks and only make minor revisions such as importing a different dataset. Given the importance of code copying in creating new D3 programs (see \autoref{sec:related-work}), we also collected $\approx20$ duplicates for each of our initial 240 notebooks (10 per vis. type), which expanded our dataset to 715 total notebooks. In the paper, we focus on the 240 unique notebooks. Since 475 of 715 notebooks are duplicates, one can extrapolate our results to the broader corpus.

Further, we define the following terms for our analysis: \textbf{decompose} refers to how users \revised{``organize," ``structure," ``break down" and otherwise} separate code within a single Observable notebook using cells~\cite{chen2023mystique}.\cut{We use this definition as an overarching term to describe how users ``organize," ``structure," and "break down" code~\cite{chen2023mystique}.}
\textbf{Modularity} refers to the extent to which users decompose their code into separate pieces, i.e., modules \revised{such as code cells} \cite{Battle2022exploring,rich2018decomposition}. \label{modular definition}\cut{ An example of a module would be an Observable code cell. } \textbf{Sources} are D3 programs that provide code for other programs. It could be a notebook on Observable (e.g. from the \href{https://observablehq.com/@d3/gallery}{D3 Gallery}) or \revised{an external} D3 program \cut{shared externally} (e.g. from \href{https://gist.github.com/}{GitHub Gist}).

\section{Can We Infer User Reasoning From D3 Code?}
\label{sec:analysis1}

We seek to understand how Observable users
reason about D3 programs and whether this can be determined indirectly through observing users' code decomposition strategies.
However, the visualization literature is unclear on how purposefully users organize reused code compared to code written from scratch.
To this end, we perform a mixed methods evaluation of our corpus to \textbf{examine} notebook code structure (\autoref{decomp stategies}), \textbf{inspect} code copying of sources (\autoref{sec:inheritance}), and \textbf{compare} decomposition strategies of corpus notebooks to their source notebooks (\autoref{decomp inheritance}).

To develop our codebook, a random sample of 15 notebooks from the six most popular D3 visualization types \revised{(observed in \cite{Bako2022streamlining, Battle2018beagle, Battle2022exploring})} were collected with the procedure in \autoref{methods}. \emph{These 15 notebooks were distinct from the 715 notebooks in the main corpus.} The \revised{two} lead authors independently annotated these notebooks, after which the entire author team met to discuss the codebook. The lead authors re-annotated the 15 notebooks again, achieving a Cohen’s Kappa inter-rater reliability score \cite{cohen1960coefficient} of 0.941. Finally, the lead authors coded the 240 unique notebooks collected from \autoref{methods} over 15 weeks. Discrepancies were resolved through discussions. In this section, we detail findings from our coding (the full codebook can be found in our supplemental materials).

\begin{figure*}
    \centering
    \includegraphics[alt={Example layouts of the three decomposition levels. Diagram shows three  notebook structures: Program-Level combines all data, geometric positional layers, functions, and scales in one cell; Chart-Level separates function-calling into a cell and then function building, scales, and data into another cell. Component-Level places each component in its own cell. Example decomposition code is shown on the left for Chart-Level and on the right for Component-Level.}, width=\linewidth]{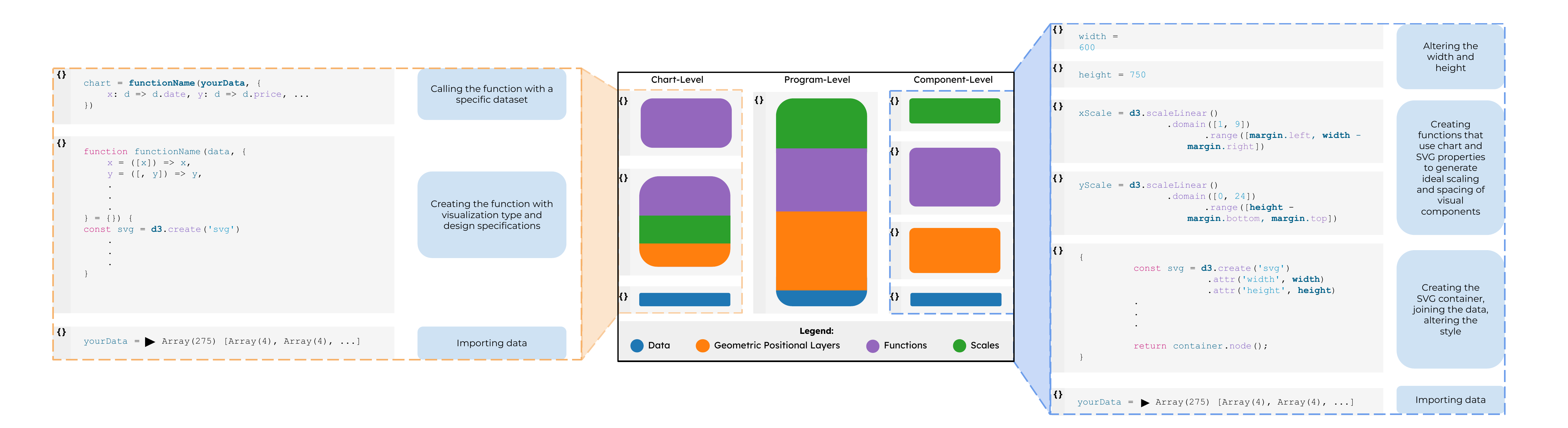}
    \vspace{-7mm}
    \caption{Centered are three abstracted notebooks with four color-coded visualization components. Program-Level has all four components in one code cell. Chart-Level has a function code cell, a function building code cell, and a data code cell. Lastly, Component-Level has all four components in different code cells. To the left is a sample of Chart-Level Decomposition. To the right is an example of Component-Level Decomposition.\vspace{-5mm}}
    \label{combinedFig}
\end{figure*}

\subsection{How Do D3 Users Decompose Programs?} \label{decomp stategies}
We examined corpus notebooks to identify the number and functionality of code cells and the relationships between cells, such as variable and function dependencies, which revealed three high-level code decomposition methods.

\textbf{Component-Level decomposition} was the most common strategy observed in our analysis, appearing in 83.8\% of notebooks. Users create each code cell as a distinct \emph{component}. Each component builds on previously defined components to implement a visualization (see \autoref{combinedFig}). Individual components also tend to fall under one step of the visualization process (e.g. importing the dataset, specifying encodings like positional scales).

\textbf{Chart-Level decomposition} was present in 7.1\% of notebooks. Illustrated in \autoref{combinedFig}, it organizes code by the target output (i.e., an entire visualization) instead of visualization steps. Users create a \emph{helper function cell} that generates the target visualization type.

\textbf{Program-Level decomposition} was present in only 6.3\% of notebooks. Code is \emph{not} decomposed and instead is placed all inside a single Observable cell. We acknowledge that users could be using white space to segment their code (see \autoref{sec:related-work}), but the low prevalence of this strategy suggests that users are deliberately choosing to forego decomposition. We interview users about their decision-making strategies when organizing D3 code in \autoref{sec:user-study}.

Only 2.9\% of notebooks show both Component- and Chart-Level structure (see supplemental materials for examples of all strategies). Overall, our findings suggest that Observable users have specific ways they prefer to organize their code.
Further, we may be able to 
\textbf{understand how users reason about their D3 code by observing how they (re)structure copied code by 
decomposition strategy (H1)}, which we explore in the following sections.

\subsection{How do D3 Users Leverage Copied Code?}
\label{sec:inheritance}

To explore our hypothesis \textbf{H1} from \autoref{decomp stategies}, we analyze code reuse in Observable notebooks. Verbatim copying may indicate that Observable users are not thinking deeply about their D3 programs. However, if we observe a range of code-copying patterns, this may be indicative of deliberate reasoning processes. 

\cut{In our codebook, }We identify four code-copying strategies; two involve no direct code reuse: \textit{``original creation,''} where users built notebooks from scratch (\revised{136 out of 240 total notebooks}, 56.7\%), and \textit{``orthogonal code forking,''} where users forked a notebook but did not reuse any \cut{of the original} code (\revised{5/240 notebooks}, 2.1\%). The remaining two strategies involve code reuse: \textit{``Observable sourced,''} where code was reused from another Observable notebook (\revised{86/240}, 35.8\%), and \textit{``outside sourced,''} where code was copied from external platforms such as GitHub (\revised{13/240}, 5.4\%). Over half of the notebooks did not copy code; of those that did, most sourced from Observable.
\revised{These percentages represent the proportion of notebooks labeled with reuse strategies (Observable- or outside-sourced) relative to the full dataset.} 
We analyze how code copying influenced selected decomposition strategies in \autoref{decomp inheritance}.

\subsection{How Does Code Copying Influence Decomposition?}
\label{decomp inheritance}

To explore \textbf{H1} further, we refine our hypothesis: if Observable notebooks are a faithful reflection of the author's reasoning process, we posit that notebook authors will (1) write code from scratch that reflects their preferred decomposition method or
(2) make substantive changes to copied code if the source clashes with how they prefer to decompose D3 code. To test this, we compared decomposition strategies in notebooks with copied code to their sources (i.e., the notebook from which the code was copied). Two key patterns emerged. First, 76.2\% of \revised{notebooks} \textit{retained Component-Level decomposition} when it was already used in the source. Second, when changing strategies, \textit{users shifted exclusively toward Component-Level decomposition}. For example, 19.8\% shifted from Chart-Level to Component-Level and 9.3\% from Program-Level to Component-Level. Further, zero notebooks shifted in the opposite direction.

\revised{13} Observable notebooks sourced code from outside \revised{Observable.}
\cut{platforms that lack Observable’s cell-based environment.}
\cut{Nonetheless, }A manual review of their code suggests that 84.62\% \revised{these} notebooks still employed Component-Level decomposition, \revised{similar to previous findings~\cite{Bako2022streamlining}.}
These findings suggest that \textbf{Component-Level decomposition is not simply a byproduct of Observable's cellular environment and may reflect how users generally reason about D3 code}.
Similarly, all 7.9\% of notebooks with inaccessible deprecated D3 Gallery notebook sources displayed Component-Level decomposition. Lastly, 80.9\% of the Original and Orthogonal notebooks used Component-Level decomposition.
Together, these results suggest that
\textbf{users prefer Component-Level decomposition, even if they do not inherit any code.}


\subsection{\textbf{Section Summary}}

Our findings show that users reason about D3 code at the Component level, whether coding from scratch or reusing copied code, suggesting that \textbf{Component-Level decomposition aligns with how users reason about D3 and supporting H1}.
\section{Validating Our Findings With User Interviews}
\label{sec:user-study}

Our analysis suggests that the structure of a D3 program could be used as a proxy for inferring how its author reasons about visualization code.
To further understand how decomposition strategies relate to how people think about their code, we conducted an IRB-approved study with N=7 D3 users about their code organization on Observable. Participants, recruited via professional networks and Observable, varied in age, education, and occupation (details in supplemental materials).

\subsection{Interview Protocol}

Participants were given an overview of the study and asked to sign a consent form. They also completed an optional survey on demographics and experience with statistics, visualization, and Observable. Interviews were conducted on Zoom, lasting an average of 34 minutes each. Interviews began with participants sharing an Observable D3 notebook, explaining its purpose, code organization, and debugging strategies. They were then asked about their D3 and Observable use, influences, and reliance on sources.

\subsection{Emerging Themes from the Interviews}

All participants shared specific reasons for their code organization. \textbf{P4} and \textbf{P6} built functions to facilitate reuse. Participants also discussed \emph{thoughtful deviations} from usual strategies. \textbf{P5}, who primarily uses Component-Level decomposition, used function calls for previously built charts to keep his visualization dashboard neat. \textbf{P4} sometimes puts all the code into a single cell to test new ideas or for single-use visualizations. \textbf{P7} uses multiple new cells when exploring new ideas and later streamlines into a single function.

Using certain decomposition levels also helped participants \emph{achieve specific goals}. \textbf{P2} mentioned his team found that using a component approach to organizing the code, instead of \revised{single cells}, led to better visualization rendering in their web application. \textbf{P3}, who used Component-Level decomposition, explained that building a visualization step-by-step helped him understand the code better. \textbf{P1} used new cells to enable interactions with his visualizations.

Debugging was a common challenge\cut{ among participants}, in part due to JavaScript's silent failures~\cite{Battle2022exploring}. \textbf{P2} and \textbf{P3} pointed to splitting code into distinct cells (i.e., Component-Level decomposition) as \textit{helping identify problems}. \textbf{P1} and \textbf{P7} typically do not use Component-Level decomposition, but created new cells to debug. This aligns with \revised{prior} work, which finds that Component-Level decomposition can make assignments easier to debug and faster to complete~\cite{charitsis2023detecting}.

Lastly, participants shared how they learned to \textit{structure their code from examples}. \textbf{P3} credits learning sources for shaping his Component-Level coding style. \textbf{P4} revised his function-writing style after finding D3 co-creator Mike Bostock's code organization to be clearer. \textbf{P6} discussed \textit{``It's easy to learn [using Observable]. I can go open anybody's notebook [and see] this is how they have written it... I can say that I learn from other people's code.''} \textbf{P6}'s code structure looks very similar to \href{https://observablehq.com/@d3/gallery}{Observable D3 Gallery} notebooks since he frequently relies on them for inspiration.

Combining these interview findings with our results from \autoref{sec:analysis1}, we argue that \textbf{users deliberately structure their D3 code to match how they reason about visualization programs, with a common preference for Component-Level decomposition.} We acknowledge that future research is needed to quantitatively validate \textbf{H1}. However, we believe these results are a strong starting point for approximating user reasoning through D3 components.

\section{What Can We Learn From Users' Decomposition Strategies in D3 Code?}\label{analysis2}

Inspired by prior work \cite{Pu2023howdata}, we reuse our results from \autoref{sec:analysis1} to compare user-made D3 components with the Grammar of Graphics (GoG) \cite{wilkinson2012grammar} and Layered GoG \cite{Wickham2010layered}, which formalize how toolkits and languages should ideally be structured.
While established formalisms often reflect how toolkit developers think, they do not necessarily reflect how end users themselves reason about visualization code~\cite{Pu2023howdata}. Thus, we seek to identify and characterize misalignments, which could reveal opportunities to improve toolkit usage and design. 
Note that we
\revised{refer to the GoG and LGoG} collectively as the GoG (Data and Aesthetic Mappings, Statistical Transformation, Geometric Object, Scale, and Coordinate System overlap with the LGoG; the GoG additionally includes Data Transformation). 

\subsection{Coverage By Visualization Type}

We calculate GoG overlap based on component counts for each visualization type in \autoref{tab:gog_coverage}. Common visualization types have greater GoG overlap. For example, Battle et al. report Geographic Map, Line Chart, Bar Chart, and Scatterplot as the most popular visualization types in their analysis of D3 usage\cut{ on the web }~\cite{Battle2018beagle}, which \cut{we observe }also have high overlap with the GoG.
\cut{However, }As complexity increases, GoG overlap drops. This suggests that \textbf{the Grammar of Graphics supports common charts but is less effective for representing customized designs, where D3 is considered more useful}~\cite{Harper2018converting, yang2024considering}, answering \textbf{RQ2}. The diminishing returns between the GoG and D3 suggest an exciting opportunity to infer higher-level abstractions from real-world toolkit usage.\cut{; in essence, a user-driven generalization of existing template-based approaches.}
Also, we emphasize that these results generalize to the full dataset of 715 notebooks, including duplicates.

\begin{table}[t]
\caption{Percent Coverage of Component by GoG Across Vis Types}
\centering
\begin{tabular}{|l|c|l|c|}
\hline
\makecell{\textbf{Visualization} \\ \textbf{Type}} & \makecell{\textbf{GoG} \\ \textbf{Coverage}} & \makecell{\textbf{Visualization} \\ \textbf{Type}} & \makecell{\textbf{GoG} \\ \textbf{Coverage}} \\
\hline
Grouped Bar Chart & 81.3\% & Box Plot & 65.3\% \\
\hline
Line Chart & 78.6\% & Hexabin & 65.2\% \\
\hline
Pie Chart & 75.6\% & Heatmap & 62.6\% \\
\hline
Bar Chart & 75.4\% & Chord & 61.2\% \\
\hline
Stacked Bar Chart & 74.2\% & Word Cloud & 59.7\% \\
\hline
Geographic Map & 74.1\% & Waffle Chart & 59.3\% \\
\hline
Scatterplot & 73.9\% & Bubble Chart & 57.1\% \\
\hline
Area Chart & 70.7\% & Sunburst & 56.8\% \\
\hline
Radial Chart & 68.3\% & Sankey & 55.8\% \\
\hline
Parallel Coord. & 66.9\% & Treemap & 54.5\% \\
\hline
Graph & 66.4\% & Voronoi & 53.9\% \\
\hline
Donut Chart & 65.8\% & Streamgraph & 48.6\% \\
\hline
\end{tabular}
\label{tab:gog_coverage}
\vspace{-5mm}
\end{table}

\subsection{Isolated Components}
We \cut{additionally }examined \cut{the }components that were \revised{often} isolated in their own cell in \autoref{isolatedcomponents}, indicating deliberate structuring decisions.\cut{ (see \autoref{decomp stategies}).}
\revised{72.9\%} of interactions are isolated, likely due to their complexity, which motivates their separation from other \cut{code} components \cite{Bako2023userdriven,Satyanarayan2017vegalite}. \cut{On the other hand,} Only 54\% of animations are separated, a likely side effect of how 
\revised{they are structured}
as calls to existing components, \revised{making them} harder to reason about and debug. These findings 
\revised{provide usage-driven support for current (e.g., \cite{Satyanarayan2017vegalite}) and future efforts}
to \revised{cover}\cut{D3 users with new grammars or toolkits that cover GoG components as well as}
overlooked areas like parameterization, animation, and interaction.

\begin{table}
[]\label{isolatedcomponents}
\caption{Percentage of \revised{isolated instances of each component}.}
\centering
\renewcommand{\arraystretch}{1.2}
\begin{tabular}{|p{4cm}|p{3cm}|}
\hline
\textbf{Component} & \textbf{Percentage Isolated} \\
\hline
Animation & 54.29\% \\
\hline
Coordinate System & 70.73\% \\
\hline
Data and Aesthetic Mappings & 93.83\% \\
\hline
Data Transformation & 87.66\% \\
\hline
Geometric Object & 74.42\% \\
\hline
Interaction & 72.90\% \\
\hline
Graph/Tree Layout & 57.14\% \\
\hline
Parameterization & 86.35\% \\
\hline
Scale & 87.50\% \\
\hline
Statistical Transformation & 61.11\% \\
\hline
\end{tabular}
\vspace{-5mm}
\end{table}

\section{Discussion: Implications for Future Research}
\textbf{Analyzing D3 Usage to Assess Existing Theory:} To the best of our knowledge, our paper provides the first qualitative analysis of overlap between D3 \revised{usage} and the GoG \cite{wilkinson2012grammar}. For \textbf{RQ2}, we find high alignment for simpler charts like bars and lines, while complex types such as word clouds and Sankey diagrams fall outside the GoG framework. Notably, \revised{usage of} essential D3 features like animations, interactions, layouts, and parameterization \revised{is} not covered by the GoG \cite{Satyanarayan2017vegalite}. While the GoG is not expected to capture every use case, our findings 
\revised{provide \emph{usage-driven evidence} for augmenting visualization grammars such as by validating recent developments in interaction grammar design (e.g., \cite{Satyanarayan2017vegalite}).}

\textbf{Leveraging Decomposition Strategies for Learning:} Our user study shows users purposefully adopt decomposition strategies to support their workflows both when coding from scratch and when reusing code, thus answering \textbf{RQ1}. Educators can leverage our findings by \textit{modularizing tutorials} or assigning Program-Level code and observing \revised{how} students restructure it as an \textit{informal assessment}. AI tools could also be trained to organize and label code by components (e.g., data handling, interactions) and provide \revised{component-focused} explanations, producing more intuitive, well-structured output.
\revised{In preliminary tests, we observed mixed results when prompting LLMs to label and generate tutorials for individual components. We observed notable hallucinations and errors, especially in the tutorials, revealing opportunities for future research.}

\textbf{Developing Resources to Enhance Visualization Design Through Component Reuse:} Component-Level decomposition can support efficient design by mapping common visualization components across multiple notebooks and visualization types. A resource linking reusable components across visualizations could help users quickly construct custom charts without starting from \revised{scratch~\cite{raghunandan2024lodestar}}. In this way, users can explore D3 code along two levels of abstraction simultaneously: Component-Level semantics and the low-level D3 syntax. Such analysis may even enable automatic component detection in other languages, helping to synthesize language abstraction levels automatically (e.g., generate Vega-Lite analogs for more complex toolkits). By providing a more structured, component-based pathway to visualization design, this method could \textit{facilitate faster design iteration and experimentation.}

\textbf{Limitations.} Notebooks were collected from Observable, representing a subset of D3 users. Some Observable users also use inaccessible private notebooks. While we employed multiple strategies to increase the rigor of our notebook and source collection (see \autoref{methods}), there are notebooks where we were unable to locate or analyze the source.
Thus, decomposition inheritance may not be fully verifiable. We encourage future studies to test our hypotheses.

\section*{Supplemental Materials}
\label{sec:supplemental_materials}
All supplemental materials are available on OSF at \revised{\url{https://osf.io/sudb8/?view_only=cc72bdc685804e478852a96297328eb8}}.
\cut{We also provide a complete copy of supplemental materials as a zip file on PCS, with a detailed \texttt{README.md} file that outlines its contents.}

\acknowledgments{
\revised{This research was funded in part by a Sloan Research Fellowship, a Mary Gates Research Scholarship, a Google gift, and NSF awards IIS-2402718, IIS-2141506, and CSGrad4US-2313998}
}

\bibliographystyle{abbrv-doi}

\bibliography{references}
\end{document}